\documentclass[twoside]{report}
\usepackage{amsmath,amsthm,amsfonts,latexsym,amssymb}

\def \bx {{\mathbf x}}

\def \NN {{\mathbb N}}

\def \CC {{\mathbb C}}
\def \oon {\frac{1}{n}}

\def \O {{\mathcal O}}

\def \HH {{\mathbb H}}

\def \B {{\mathcal B}}

\def \S {{\mathcal S}}

\def \F {{\mathcal F}}
\def \M {{\mathcal M}}
\def \TR {\mathrm{Tr}}

\def \NN {{\mathbb N}}

\def \CC {{\mathbb C}}

\def \bow {{\mathbf 0}}

\outer\def\proof{\smallbreak\noindent{\bf Proof.}\enspace}

\newtheorem{thm}{Theorem}[section]
\newtheorem{lem}[thm]{Lemma}
\newtheorem{cor}[thm]{Corollary}
\newtheorem{dfn}[thm]{Definition}

\newtheorem{prp}[thm]{Proposition}

\begin{document}
\chapter{Classical and Quantum Propagation of Chaos}
\author{Alex D. Gottlieb}

\pagestyle{myheadings} \markboth{Alex Gottlieb}{Propagation of
Chaos in Classical and Quantum Kinetics}

\section{Overview}

   The concept of molecular chaos dates back to Boltzmann~\cite{Boltzmann}, who derived
the fundamental equation of the kinetic theory of gases under the
hypothesis
 that the molecules of a
nonequilibrium gas are in a state of ``molecular disorder.'' The
concept of {\it propagation} of molecular chaos is due to Kac
\cite{Kac55, Kac}, who called it ``propagation of the Boltzmann
property'' and used it to derive the homogeneous Boltzmann
equation in the infinite-particle limit of certain Markovian gas
models (see also \cite{Gru, Szn}). McKean \cite{McK66,McK68}
proved the propagation of chaos for systems of interacting
diffusions that yield diffusive Vlasov equations in the mean-field
limit.   Spohn \cite{Spohn} used a quantum analog of the
propagation of chaos to derive time-dependent Hartree equations
for mean-field Hamiltonians, and his work was extended in
\cite{Alicki} to open quantum mean-field systems.

This article examines the relationship between classical and
quantum propagation of chaos.   The rest of this introduction
reviews some ideas of quantum probability and dynamics.
Section~\ref{Sect2} discusses the classical and quantum concepts
of propagation of chaos.  In Section~\ref{Sect3}, {\it classical}
propagation of chaos is shown to occur when {\it quantum} systems
that propagate quantum molecular chaos are suitably prepared,
allowed to evolve without interference, and then observed.   Our
main result is Corollary~\ref{interesting}, which may be
paraphrased as follows:

 Let $\O$ be a complete observable of a single particle,
taking its values in a countable set $J$, and let $\O_i$ denote
the observable $\O$ of particle $i$ in a system of $n$
distinguishable particles of the same species.  Suppose we allow
that quantum $n$-particle system to evolve freely, except that we
periodically measure $\O_1,\O_2,\ldots,\O_n$.   The resulting time
series of measurements is a Markov chain in $J^n$. {\it If the
sequence of $n$-particle dynamics propagates quantum
 molecular chaos, then these derived Markov chains propagate chaos in the classical sense}.

Results like this may be of interest to probabilists who already
know some examples of the propagation of chaos and who may be
surprised to learn of novel examples arising in quantum dynamics.
An effort has been made here to expound the propagation of quantum
molecular chaos for such an audience, while the classical
propagation of chaos {\it per se} is discussed only briefly in
Section~\ref{CMC}. The reader is referred to \cite{Sznitman} and
\cite{Meleard} for two definitive surveys of the classical
propagation of chaos.

\subsection{Quantum kinematics}

In quantum theory, the state of a physical system is inherently
statistical: the state of a system $\S$ does not {\it determine}
whether or not $\S$ has a given property, but rather, the state
provides only the {\it probability} that $\S$ would be found to
have that property, if we were to check for it.  The properties
that $\S$ might or might not have are represented by orthogonal
projectors on some Hilbert space.      If a projector $P$
represents a property P of $\S$, then the complementary property
NOT P is represented by $I-P$.  The identity and zero operators
$I$ and $\bow$ represent the trivial properties TRUE and FALSE
respectively. If P and Q are properties whose orthogonal
projectors are $P$ and $Q$, the properties (P AND Q) and (P OR Q)
are defined if and only if $P$ and $Q$ commute, in which case $PQ
= QP$ represents (P AND Q) and $P + Q - PQ$ represents (P OR Q). A
countable {\it resolution of the identity} is a countable family
of projectors $\{P_j\}$ such that
\[
     P_j P_{j'} \ = \ P_{j'}P_j \ = \ \bow \ ; \qquad \forall j \ne j'
\]
and $I = \sum_j P_j$.  This represents a partition of the space of
outcomes of a measurement on $\S$ into a countable set of
elemental properties, which are mutually exclusive and
collectively exhaustive.   A {\it state} $\omega$ of the system
$\S$ is a function that assigns probabilities to the properties of
$\S$, or their projections. Thus we suppose that $\omega(I)=1,\
\omega(\bow)=0$ and also that
\[
\omega(I-P)+\omega(P) \ = \ \omega(I) \ = \ 1.
\]
Indeed, it is rational to suppose that any state $\omega$ must
satisfy
\begin{equation}
\label{stateEquation}
        1 \ = \ \omega(I) \ = \ \omega\big(\sum_j P_j\big) \ =
        \ \sum_j \omega(P_j)
\end{equation}
for any countable resolution of the identity $\{P_j\}$.

Having made these introductory comments, we revert to a more
technical description of the mathematical set-up. Suppose the
properties of a quantum system $\S$ are represented by orthogonal
projectors in $\B(\HH)$, the bounded operators on a Hilbert space
$\HH$.  The {\it statistical states} (also called simply {\it
states}) of that quantum system are identified with the normal
positive linear functionals on $\B(\HH)$ that assign $1$ to the
identity operator. A positive linear functional $\omega$ on
$\B(\HH)$ is {\it normal} if
\[
      \sum_{a \in A} \omega(P_a)  = 1
\]
whenever $\{P_a\}_{a \in A}$ is a family of commuting projectors
that sum to the identity operator (i.e., the net of finite partial
sums of the projectors converges in the weak operator topology to
the identity). Normal states are in one-one correspondence with
{\it density operators}, positive trace-class operators of trace
$1$.  If $D$ is a density operator on $\HH$ then $A \mapsto
\TR(DA)$ defines a normal state on $\B(\HH)$; conversely, every
normal state $\omega$ on $\B(\HH)$ is of the form $\omega(A) =
\TR(DA)$ for some density operator $D$.   The density operators
form a closed convex subset of the trace-class operators, which is
a Banach space with $\|T\| = \TR(|T|)$. The dual of the Banach
space of trace class operators is $\B(\HH)$ with its operator
norm.

 In the Heisenberg picture of quantum dynamics
--- where the state is constant while the operators corresponding
to observables change --- the dynamics are given by unitarily
implemented automorphisms of the bounded operators on a Hilbert
space.  That is, for each $\tau \ge 0$ there exists a unitary
operator $U(\tau)$ such that a property represented by $P$ at time
$t=0$ is represented by
\begin{equation}
\label{Heisenberg}
       P(\tau) \ = \ U(\tau)^*P U(\tau)
\end{equation}
at time $t=\tau$.  In the Schr\"odinger picture of dynamics, the
density operator $D$ of the quantum state changes in time while
the projectors $P$ that represent properties of the system remain
fixed. The Schr\"odinger formulation of the dynamics corresponding
to (\ref{Heisenberg}) is
\[
        D(\tau) \ = \  U(\tau) D(0)  U(\tau)^*,
\]
because, for any $P \in \B(\HH)$,
\[
    \TR [D P(\tau)]  \ = \   \TR [D \ U(\tau)^*P U(\tau)] \ = \ \TR [U(\tau) D  U(\tau)^* P]
    \ = \ \  \TR [D(\tau) P].
\]
The Heisenberg picture of dynamics suggests a generalization where
the dynamics $A \mapsto U^*AU$ are replaced by more general
endomorphisms of $\B(\HH)$, namely, by completely positive maps
$\phi: \B(\HH) \longrightarrow \B(\HH)$ such that $\phi(I)=I$.
 These
endomorphisms include the automorphisms $A \mapsto U^*AU$ of the
Heisenberg picture of quantum dynamics, but also describe maps of
observables $A \mapsto \phi(A)$ effected by the intervention of
measurements, randomization, and coupling to other systems.    A
map $\phi$ is positive if it maps nonnegative operators to
nonnegative operators, and it is {\it completely positive} if $
\phi\otimes \hbox{id}_d $ is positive whenever $\hbox{id}_d$ is
the identity on $\B(\CC^d)$ for any finite $d$. Requiring $\phi$
to be positive and unital (unit preserving) is necessary to ensure
that (\ref{stateEquation}) holds, at least for any finite
resolution of the identity.  The {\it complete} positivity of
$\phi$ ensures the positivity of the dynamics of certain
extensions of the original system $\S$, where $\S$ is considered
together with a physically independent, finite-dimensional quantum
system.   To pass to the Schr\"odinger picture we must impose the
further technical requirement that the map $\phi$ be {\it normal},
i.e., $\phi$ is assumed to be such that
\[
           \lim \phi (A_{\alpha}) \ = \ \phi(A)
\]
whenever $\{A_{\alpha}\}$ is a monotone increasing net of positive
operators with least upper bound $A$.  This way the Schr\"odinger
dynamics of the normal state can be defined as the ``predual" of
the Heisenberg dynamics; if $\phi$ is normal then the relation
\[
    \TR(\phi_*(D) A) = \TR(D \phi(A) )  \qquad \forall A \in \B(\HH)
\]
implicitly defines a trace-preserving map $\phi_*$ known as the
{\it predual of} $\phi$.    In the Schr\"odinger picture, the
density operator $D$ that describes the quantum state undergoes
the transformation $D \mapsto \phi_*(D)$, where $\phi$ is a normal
completely positive unital endomorphism of $\B(\HH)$.

The description of a quantum system evolving continuously in time
requires a normal and completely positive endomorphism of
$\B(\HH)$ for each $t>0$, to describe the change of observables
(in the Heisenberg picture) from time $0$ to time $t$.  A {\it
quantum dynamical semigroup}, or QDS, is a family $ \{\phi_t\}_{t
\ge 0} $
 of normal completely positive (and unital) endomorphisms
of the bounded operators on some Hilbert space $\HH$, which is a
semigroup (i.e., $\phi_0 = \hbox{id}$ and $\phi_t \circ \phi_s =
\phi_{t+s}$ for $s,t\ge 0$) and which has weak*-continuous
trajectories: for any $B \in \B(\HH)$ and any trace class operator
$T$
\[
            \TR(T \ \phi_t(B))
\]
is continuous in $t$.   (We will not need the continuity of
trajectories in this article.) Quantum dynamical semigroups
describe the continuous change of the state of an open quantum
system whose dynamics are autonomous and Markovian.  Many models
of open quantum systems are QDSs (but not all, viz. \cite{Kohen}).
We will use the notation $(\phi)_t$ for the whole QDS:
\begin{equation}
\label{QDS}
      (\phi)_t \ = \  \{\phi_t\}_{t \ge 0}.
\end{equation}

\section{Classical and Quantum Molecular Chaos}
\label{Sect2}

\subsection{Classical molecular chaos}
\label{CMC}

Molecular chaos is a type of stochastic independence of particles
manifesting itself in an infinite-particle limit.

Let $\Omega^n$ be the $n$-fold Cartesian power of a measurable
space $\Omega$.  A probability measure $p$ on $\Omega^n$ is called
{\it symmetric} if
\[
    p(E_1 \times E_2 \times \cdots \times E_n) =
    p(E_{\pi(1)} \times E_{\pi(2)} \times \cdots \times E_{\pi(n)})
\]
for all measurable sets $E_1, \ldots, E_n \subset \Omega$ and all
permutations $\pi$ of $\{1,2,\ldots, n\}$.  For $k \le n$, the
$k$-{\it marginal} of $p$, denoted $p^{(k)}$, is the probability
measure on $S^k$ satisfying
\[
   p^{(k)}( E_1 \times E_2 \times \cdots \times E_k) =
    p( E_1 \times \cdots \times E_k \times \Omega \times \cdots \times \Omega)
\]
for all measurable sets $E_1, \ldots, E_k \subset \Omega$. In the
context of classical probability theory, one defines molecular
chaos as follows \cite{Sznitman}:

\begin{dfn}
\label{ClassicalMolecularChaos} {\rm
 Let $\Omega$ be a separable metric space. Let $p$ be a probability
measure on $\Omega$, and for each $n
 \in \NN$, let $p_n$ be a symmetric probability measure on
$\Omega^n$.

The sequence $\{p_n\}$ is $p$-{\bf chaotic} if the $k$-marginals
$p_n^{(k)}$ converge weakly to $p^{\otimes k}$ as $n
\longrightarrow \infty$, for each fixed $k \in \NN$. }
\end{dfn}

A sequence, indexed by $n$, of $n$-particle dynamics {\it
propagates chaos} if molecularly chaotic sequences of initial
distributions remain molecularly chaotic for all time under the
$n$-particle dynamical evolutions. In the classical contexts
\cite{Kac,McK66,Meleard,Sznitman} the dynamics are Markovian and
the state spaces are usually taken to be separable and metrizable.
Accordingly, in my dissertation \cite{Thesis} I defined
propagation of chaos in terms of Markov transition kernels, as
follows:
\begin{dfn}[Classical Propagation of Chaos]
\label{PropagationOfClassicalChaos} {\rm Let $\Omega$ be a
separable metric space.  For each $n \in \NN$, let $K_n:\Omega^n
\times \sigma(\Omega^n) \longrightarrow [0,1]$ be a Markov
transition kernel which is invariant under permutations in the
sense that
\[
     K_n(\bx,E) = K_n(\pi \cdot \bx, \pi \cdot E)
\]
for all permutations $\pi$ of the $n$ coordinates of $\bx$ and the
points of $E \subset \Omega^n$.  Here, $\sigma(\Omega^n)$ denotes
the Borel $\sigma$-field of $\Omega^n$.

The sequence $\{K_n\}_{n=1}^{\infty}$ {\bf propagates chaos} if
the molecular chaos of a sequence  $\{ p_n \}$ entails the
molecular chaos of the sequence
\begin{equation}
\label{induced} \left\{ \int_{\Omega^n} K_n(\bx,\cdot) p_n (d\bx)
\right\}_{n=1}^{\infty} \ .
\end{equation}
}
\end{dfn}

The preceding formulation of the propagation of molecular chaos is
technically straightforward but not flexible enough to cover the
weaker kinds of propagation of chaos phenomena that occur in
several applications, most notably in the landmark derivation of
the Boltzmann equation due to Lanford and King \cite{Lanford}.
Nonetheless, it is still worthwhile to make
Definition~\ref{PropagationOfClassicalChaos}.  Those models that
exhibit weak propagation of chaos phenomena usually have less
realistic regularizations that propagate molecular chaos in the
sense of Definition~\ref{PropagationOfClassicalChaos}.  Moreover,
this definition has the pleasant feature that it implies that
$\{K_n \circ L_n\}$ propagates molecular chaos when $\{K_n\}$ and
$\{L_n\}$ do.

\subsection{Quantum molecular chaos}

 The
Hilbert space of pure states of a collection of $n$
distinguishable components is $\HH_1 \otimes \cdots \otimes
\HH_n$, where $\HH_i$ is the Hilbert space for the $i^{th}$
component. The Hilbert space for $n$ distinguishable components of
the same species will be denoted $\HH^{\otimes n}$. If $D_n$ is a
density operator on $\HH^{\otimes n}$, then its $k$-marginal, or
partial trace, is a density operator on $\HH^{\otimes k}$ that
gives the statistical state of the first $k$ particles.  The
$k$-marginal may be denoted $ \TR^{(n-k)}D_n$ and defined as
follows: Let $O$ be any orthonormal basis of $\HH$.  If $x \in
\HH^{\otimes k}$ with $k < n$ then for any $w,x \in \HH^{\otimes
k}$
\[
\begin{array}{l}
   \left< \TR^{(n-k)} D_n(w),x \right> \ =   \\
                                             \\
\qquad         \sum\limits_{y_1,\ldots,y_{n-k} \in O} \left< D_n(
w \otimes
   y_1 \otimes \cdots \otimes y_{n-k}), x \otimes y_1 \otimes \cdots \otimes y_{n-k}   \right> . \\
\end{array}
\]
A linear functional $\omega$ on $\B(\HH^{\otimes n})$ is {\it
symmetric} if it satisfies
\[  \omega(A_1 \otimes \cdots \otimes A_n) = \omega(A_{\pi(1) }
     \otimes A_{\pi(2)} \otimes \cdots \otimes A_{\pi(n)})
\]
for all permutations $\pi$ of $\{1,2,\ldots,n \}$ and all
$A_1,\ldots,A_n \in \B(\HH)$. For each permutation $\pi$ of
$\{1,2,\ldots,n \}$, define the unitary operator $U_{\pi}$ on
$\HH^{\otimes n}$ whose action on simple tensors is
\begin{equation}
\label{Upi}
     U_{\pi}(x_1 \otimes x_2 \otimes \cdots \otimes x_n) = x_{\pi(1)}
\otimes x_{\pi( 2)}\otimes \cdots \otimes x_{\pi (n)}.
\end{equation}
A density operator $D_n$ represents a symmetric functional on
$\B(\HH^{\otimes n})$ if and only $D_n$ commutes with each
$U_{\pi}$.   Two special types of symmetric density operators are
Fermi-Dirac densities, which represent the statistical states of
systems of fermions, and Bose-Einstein densities, which represent
the statistical states of systems of bosons. Bose-Einstein density
operators are characterized by the condition that $D_n U_{\pi} =
D_n$ for all permutations $\pi$, and Fermi-Dirac densities are
characterized by the condition that $D_n U_{\pi} =
\hbox{sign}(\pi) D_n$ for all $\pi$.

To recapitulate, $n$-component states are given by density
operators on $\HH^{\otimes n}$ and, in the Sch\"odinger picture,
the dynamics transforms an initial state $A \mapsto \TR(DA)$ into
a state of the form $A \mapsto \TR(D \phi(A))$, where $\phi$ is a
normal completely positive unital endomorphism of $\B(\HH^{\otimes
n})$. This is the context of the following two definitions:

\begin{dfn}
\label{NoncommutativeChaosNormal} {\rm
  Let $D$ be a density operator on $\HH$, and for each
$n \in \NN$, let $D_n$ be a symmetric density operator on  $
\HH^{\otimes n}$.

The sequence $\{D_n\}$ is $D$-{\bf chaotic in the quantum sense}
if, for each fixed  $k \in \NN$, the density operators
$\TR^{(n-k)}D_n$ converge in trace norm to $D^{\otimes k}$ as $n
\longrightarrow \infty$.

The sequence $\{D_n\}$ is {\bf quantum molecularly chaotic} if it
is $D$-chaotic in the quantum sense for some density operator $D$
on $\HH$. }
\end{dfn}

The definitions of classical and quantum molecular chaos are
somewhat incongruous.  This definition of quantum molecular chaos
requires that the marginals converge in the trace norm, whereas
the notion of classical molecular chaos used in Probability Theory
requires {\it weak} convergence of the marginals.  In fact,
Definition~\ref{ClassicalMolecularChaos} of classical molecular
chaos and the ``commutative" version of
Definition~\ref{NoncommutativeChaosNormal} (obtained by extending
that definition of quantum molecular chaos to commutative von
Neumann algebras) are not equivalent! Nonetheless, I have chosen
Definition~\ref{NoncommutativeChaosNormal} because the attractive
theory of quantum mean-field kinetics presented in
Section~\ref{Spohn'sSection} favors a formulation of quantum
molecular chaos in terms of the trace norm.

\begin{dfn}[Propagation of Quantum Molecular Chaos]
\label{PropagationOfQuantumChaos} {\rm For each $n \in \NN$, let
$\phi_n$ be a normal completely positive map from $\HH^{\otimes
n}$ to itself that fixes the identity and which commutes with
permutations, i.e., such that
\begin{equation}
\label{PermutationsPhi}
    \phi_n(U^*_{\pi} A U_{\pi}) = U^*_{\pi} \phi_n (A)U_{\pi}
\end{equation}
for all $A \in \B(\HH^{\otimes n})$ and all permutations $\pi$ of
$\{1,2,\ldots,n\}$, where $U_{\pi}$ is as defined in (\ref{Upi}).

The sequence $\{\phi_n\}$ {\bf propagates quantum molecular chaos}
if the quantum molecular chaos of a sequence of density operators
$\{ D_n \}$ entails the quantum molecular chaos of the sequence
$\{ \phi_{n*}(D_n) \}$. }
\end{dfn}

We shall soon find that there are interesting examples of quantum
dynamical semigroups $(\phi_n)_t$ with the collective property
that $\{\phi_{n,t}\}$ propagates quantum molecular chaos for each
fixed $t > 0$.   When this happens, it is convenient to say that
the sequence $\{(\phi_n)_t\}$ of QDSs propagates chaos.
\begin{dfn}
\label{PoCforQDS}
 For
each $n$ let $(\phi_n)_t$ be a QDS on $\B(\HH^{\otimes n})$ that
satisfies the permutation condition (\ref{PermutationsPhi}). The
sequence $\{(\phi_n)_t\}$ {\bf propagates molecular chaos} if
$\{\phi_{n,t}\}$ propagates quantum molecular chaos for every
fixed $t > 0$.
\end{dfn}

\subsection{Spohn's quantum mean-field dynamics}
\label{Spohn'sSection}

 There are several
successful mathematical treatments of quantum mean-field dynamics.
One of them, due to H. Spohn, relies upon the concept of
propagation of quantum molecular chaos. Spohn's theorem
\cite{Spohn} constitutes a rigorous derivation of the
time-dependent Hartree equation for bounded mean-field potentials.

Let $V$ be a bounded Hermitian operator on $\HH \otimes \HH $ such
that $V U_{(12)} = U_{(12)} V(y \otimes x)$, representing a
symmetric two-body potential. Let $V_{1,2}$ denote the operator on
$ \HH^{\otimes n}$ defined by
\begin{equation}
\label{Vnij}
   V_{1,2}(x_1\otimes x_2 \otimes \cdots \otimes x_n) = V(x_1 \otimes
             x_2) \otimes x_3 \otimes \cdots \otimes x_n ,
\end{equation}
and for each $i ,j \le n$ with $i < j$, define $V_{ij}$ similarly,
so that it acts on the $i^{th}$ and $j^{th}$ factors of each
simple tensor. This may be accomplished by setting $ V_{ij} =
U^{*}_{\pi}V_{1,2}  U_{\pi}$, where $\pi = ( 2 j)(1 i )$ is a
permutation that puts $i$ in the first place and $j$ in the second
place, and $U_{\pi}$ is as defined in (\ref{Upi}). Define the
$n$-particle Hamiltonians $H_n$ as the sum of the pair potentials
$V_{ij}$, with common coupling constant $1/n$:
\begin{equation}
\label{mean-field Ham}
   H_n = \oon \sum_{i < j} V_{ij}.
\end{equation}
If $D_n$ is a state on $ \HH^{\otimes n}$, let $D_n(t)$ denote the
state of an $n$-particle system that was initially in state $D_n$
and which has undergone $t$ units of the temporal evolution
governed by the Hamiltonian (\ref{mean-field Ham}):
\begin{equation}
\label{evolved state}
     D_n(t) =  e^{-i H_n t /\hbar} D_n e^{i H_n t /\hbar} .
\end{equation}

\begin{thm}[Spohn]
\label{Spohn'sTheorem} Suppose $D$ is a density operator on $\HH$
and $\{ D_n \}$ is a $D$-chaotic sequence of symmetric density
operators on $\HH^{\otimes n}$. Then the sequence of density
operators $\{ D_n(t) \}$ defined in (\ref{mean-field Ham}) and
(\ref{evolved state}) is $D(t)$-chaotic, where $D(t)$ is the
solution at time $t$ of the following initial-value problem in the
Banach space of trace-class operators:
\begin{eqnarray}
       \frac{d}{dt} D(t) & = &
            -\frac{i}{\hbar}\TR^{(n-1)}[V,D(t)\otimes D(t)]
                                       \nonumber \\
       D(0) & = & D.             \label{VlasovEq}
\end{eqnarray}

\end{thm}

In other words, if $H_n$ is as in (\ref{mean-field Ham}) and
$
     \phi_n(A) \ = \  e^{i H_n t /\hbar}\ A \  e^{-i H_n t /\hbar}
$
 then the sequence $\{\phi_n\}$ propagates quantum molecular
 chaos.  See \cite{Spohn} for a short proof.

It must be emphasized that the preceding approach does not apply
to systems of Fermions.   In other words,
Theorem~\ref{Spohn'sTheorem} yields time-dependent Hartree
equations but not time-dependent Hartree-{\it Fock} equations
(which include an exchange term that enforces the Pauli
Principle).   The problem is that there exists no molecularly
chaotic sequence of Fermi-Dirac states because the antisymmetry of
Fermi-Dirac states is incompatible with the factorization of
molecularly chaotic states.   To derive time-dependent
Hartree-Fock equations despite this problem is a goal of current
research \cite{Norbert}.

Spohn's approach can be generalized to handle Hamiltonians which
involve the usual unbounded kinetic energy operators, and to
handle {\it open} quantum mean-field systems.   In \cite{Alicki},
Theorem~\ref{Spohn'sTheorem} is extended to systems where the free
motion of the single particle may have an unbounded self-adjoint
generator and the two-particle generator may have the Lindblad
form.  The ``propagation of quantum molecular chaos" is called the
``mean-field property" in that article.

\section{Classical Manifestations of the Propagation of Quantum Molecular Chaos}
\label{Sect3}

  A sequence $\{\phi_n\}$ that propagates quantum molecular
chaos mediates a variety of instances of the propagation of
classical molecular chaos.

Given a $D$-chaotic sequence of density operators $\{D_n\}$ one
can produce a variety of molecularly chaotic sequences $\{q_n\}$
of probability measures.   For each single-particle measurement
$\M$, the joint probabilities $q_n$ of the outcome of applying
$\M$ to all of the particles form a molecularly chaotic sequence.
Conversely, there are ways to convert a $p$-chaotic sequence
$\{p_n\}$ of probability measures into a $D$-chaotic sequence of
density operators $\{D_n\}$.   Suppose we are presented with a
sequence of quantum dynamics $\{\phi_n\}$ that propagates quantum
molecular chaos.   We can first {\it encode} a sequence of
molecularly chaotic probabilities $\{p_n\}$ as a quantum
molecularly chaotic sequence of density operators $\{D_n\}$, then
allow $\{D_n\}$ to {\it develop} into a new sequence
$\{\phi_{n*}(D_n)\}$ under the dynamics, and finally {\it read}
the resulting quantum states by applying some single-particle
measurement to each of the particles.  This procedure converts one
molecularly-chaotic sequence of probability measures into another,
i.e., it propagates chaos in the classical sense.

The next three sections examine the encode/develop/read-procedure
that converts quantum molecular chaos to classical molecular
chaos. Finally, in Section~\ref{complete observables}, we show how
to produce Markov chains that propagate chaos by observing quantum
processes that propagate chaos.

\subsection{Generalized measurements and the reading procedure}
\label{generalized measurements}

 Let $\HH$ be a Hilbert space and $(\Omega,\F)$ a measurable
space. A {\it positive operator valued measure}, or POVM, is a
function $X(E)$ from $\F$ to the positive operators on $\HH$ which
is countably additive with respect to the weak operator topology:
\
\[
      \sum_{i = 1}^{\infty} X(E_i) \ = \ X(E)
\]
in the weak operator topology whenever the sets $E_i \in  \F$ are
disjoint and $E = \cup E_i$.  In the special case that the
positive operators $X(E)$ are self-adjoint projections such that
$X(E)X(F) = \bow $ when $E\cap F = \emptyset$, the POVM is just a
resolution of identity on $(\Omega,\F)$. A POVM defines an affine
map $D \mapsto p_{D,X}$ from density operators on $\HH$ to
probability measures on $\Omega$.  For any density operator $D$,
\begin{equation}
\label{genericLabel}
        p_{D,X}(\cdot) \ = \ \TR\left[DX(\cdot)\right]
\end{equation}
is a countably additive probability measure on $(\Omega,\F)$.

POVMs correspond to {\it generalized} $\Omega$-valued measurements
just as the spectral decomposition of a self-adjoint operator
corresponds to a simple measurement.   A generalized measurement
is realized by performing a simple measurement on a composite
system consisting of the system of interest an ``ancilliary
system" that has been prepared to have state $E$, without allowing
any interaction between the system itself and the ancilliary
system or the environment.  Suppose that $\HH_a$ is the Hilbert
space of an ancilliary space that has been prepared in state $E$,
and $P(d\omega), \omega \in \Omega$ is the spectral measure
belonging to an observable $\O$ on the composite system
$\HH\otimes \HH_a$. If the system of interest is in state $D$ when
$\O$ is measured, a random $\omega \in \Omega$ is produced,
governed by the probability law $\TR [(D\otimes E)\ P(d\omega)]$.
This probability law has the form (\ref{genericLabel}), since
\[
    \TR [(D\otimes E)\ P(d\omega)] \ = \
    \TR [D \ \TR^{(1)}((I\otimes E)P(d\omega))] \ = \
    \TR(DX(d\omega))
\]
where $X(d\omega)$ is the POVM
\[
      X(d\omega) \ = \ \TR^{(1)}((I\otimes E^{1/2})P(d\omega)(I\otimes
      E^{1/2})).
\]
This shows that the outcome of a generalized measurement is
governed by some POVM as in (\ref{genericLabel}).   Conversely, it
can be shown that any POVM arises in this way from some
conceivable (but perhaps impracticable) generalized measurement of
the type we have just described \cite{Holevo}.

The following lemma has a straightforward proof, which we omit.
\begin{lem}
\label{propone}
    Let $\Omega$ be a separable metric space with Borel $\sigma$-field $\sigma(\Omega)$, and let
$X:\sigma(\Omega) \longrightarrow \B(\HH)$ be a POVM on $\Omega$.

For each $n$, let $D_n$ be a symmetric density operator on
$\B(\HH^{\otimes n})$, and suppose that $\{D_n\}$ is $D$-chaotic
in the quantum sense.

Then the sequence of probability measures $\{p_n\}$ is
$p$-chaotic, $p_n$ and $p$ being defined by
\begin{eqnarray*}
    p(A) & = & \TR(D A)  \\
     p_n(A_1\times A_2 \times \cdots \times A_n)  & = &
      \TR\left(   D_n \ X(A_1)\otimes \cdots \otimes X(A_n)\right) .
    \\
\end{eqnarray*}

\end{lem}

\subsection{Lemmas concerning the encoding procedure}
\label{encode}

The procedure for encoding probability measures as density
operators depends on our choice of a density operator valued
function $D(\omega)$ on the single-particle space $\Omega$, now
assumed to be a separable metric space. Let us choose a function
$D(\omega)$ from $\Omega$ to the density operators on a Hilbert
space $\HH$, and assume $D$ is continuous for technical
convenience. Then we can convert a probability measure $p_n$ on
$\Omega^n$ into the density operator
\[
       \int_{\Omega^n}   D(\omega_1)\otimes D(\omega_2)\otimes\cdots \otimes D(\omega_n)
       \  p_n(d\omega_1 d\omega_2 \cdots
       d\omega_n)
\]
on $\HH^{\otimes n}$.  If $\{p_n\}$ is a $p$-chaotic sequence of
symmetric measures on $\Omega^n$ then the corresponding sequence
of density operators is quantum molecularly chaotic --- but only
in a weak sense!  Think of the density operators as a subset of
the Banach space of trace-class operators.  Each continuous linear
functional on this space has the form
\[
             T \ \longmapsto \ \TR (TB)
\]
where $B$ is a bounded operator, for $\B(\HH)$ is the Banach dual
of the space of trace-class operators on $\HH$.  A sequence
$\{T_n\}$ of trace-class operators on $\HH$ is {\it weakly
convergent} if
\[
         \lim_{n \rightarrow \infty}  \TR (T_n B) \ = \ \TR (TB)
\]
for all $B \in \B(\HH)$.

\begin{lem}
\label{lemma0}
    Let $\Omega$ be a separable metric space and suppose $\{p_n\}$ is a $p$-chaotic sequence of
symmetric measures on $\Omega^n$. Let $D(s)$ be a continuous
function from $\Omega$ to the density operators on a Hilbert space
$\HH$.

 Define $\bar{D}$ and $D_n$ by
\begin{eqnarray}
\label{meandensity}
      \bar{D}  & = &  \int_{\Omega} D(\omega) p(d\omega)  \nonumber \\
     D_n & = & \int_{\Omega^n}   D(\omega_1)\otimes D(\omega_2)\otimes\cdots \otimes D(\omega_n)
            \  p_n(d\omega_1d\omega_2\cdots d\omega_n) .
     \nonumber  \\
\end{eqnarray}
Then, for each $k$, the sequence of marginals $\{\TR^{(n-k)}D_n
\}$ converges weakly to $\bar{D}^{\otimes k}$ as $n
\longrightarrow \infty$.
\end{lem}

\proof The integrals in (\ref{meandensity}) may be defined as
Bochner integrals in the Banach space of trace-class operators
(see \cite{Hille} Theorem~$3.7.4$).  The partial trace is a
bounded operator from the Banach space of trace-class operators on
$\HH^n$ to the Banach space of trace-class operators on $\HH^k$,
so Theorem~$3.7.12$ of \cite{Hille} implies that
\begin{eqnarray}
\label{pfeq1}
  &   &
      \TR^{(n-k)} \int_{\Omega^n}   D(\omega_1)\otimes\cdots \otimes D(\omega_n)
       \  p_n(d\omega_1d\omega_2\cdots d\omega_n)
      \nonumber \\
  &   &   \qquad \quad \ = \ \ \
     \int_{\Omega^n}  D(\omega_1)\otimes\cdots \otimes D(\omega_k) \ p_n(d\omega_1 d\omega_2\cdots d\omega_n) .
     \nonumber  \\
\end{eqnarray}
Since Bochner integration commutes with application of bounded
linear functionals, the right hand side of (\ref{pfeq1}) converges
weakly to
\[
     \int_{\Omega^k}  D(\omega_1)\otimes D(\omega_2)\otimes\cdots \otimes
     D(\omega_k)\
     p^{\otimes k}(d\omega_1 d\omega_2 \cdots d\omega_k)
     \ = \
     \bar{D}^{\otimes k}
\]
as $n \longrightarrow \infty$. \hfill $\square$

\medskip

The preceding lemma does not conclude that $\{D_n \}$ is quantum
molecularly chaotic in the sense of our
Definition~\ref{NoncommutativeChaosNormal}, which requires
convergence of the partial traces in trace norm.   Some additional
conditions seem necessary in order to conclude that the encoding
procedure produces quantum molecular chaos. The easiest thing to
do is suppose that the quantum systems involved are finite
dimensional. This affords a quick way to construct Markov
transitions on any space $\Omega$. The opposite approach is to let
the mediating quantum dynamics occur in any Hilbert space $\HH$
but to suppose that $\Omega$ is discrete.  We follow these two
approaches in the next two lemmas:

\begin{lem}
\label{lemma1}
 If $\HH$ is a finite dimensional Hilbert space
$\CC^d$, and $D_n$ and $D$ are as in the statement of
Lemma~\ref{lemma0}, then the sequence of states $\{D_n\}$ is
$\bar{D}$-chaotic.
\end{lem}

\proof
From Lemma~\ref{lemma0} we know that $\{\TR^{(n-k)}D_n \}$
converges weakly to $\bar{D}^{\otimes k}$
 as $n \longrightarrow \infty$. But a sequence of trace-class operators on $\CC^d$
converges in trace norm if it converges weakly, since the Banach
space of trace class operators on on $\CC^d$ is
finite-dimensional.  Hence, $\lim\limits_{n \rightarrow \infty}
    \TR^{(n-k)}(D_n) = \bar{D}^{\otimes k}$ in trace norm, as required.
\hfill $\square$

Now we consider a set $J$ with a discrete topology and
$\sigma$-field.  We note that if the word ``separable" were
removed from Definition~\ref{ClassicalMolecularChaos} of classical
molecular chaos then the following lemma would hold even for
uncountable sets $J$:

\begin{lem}
\label{lemma2}

Let $J$ be a countable set equipped with the discrete topology and
its Borel $\sigma$-field (so that every subset of $J$ is
measurable), and let $\{D(j)\}_{j \in J}$ be a family of density
operators on $\HH$ indexed by $J$.

 Suppose $p$ is a probability measure on $J$ and $\{p_n\}$ is a
$p$-chaotic sequence of probability measures. Define
\begin{eqnarray}
       \bar{D}   & = &
         \sum\limits_{j \in J}
            p(j) D(j)   \nonumber  \\
     D_n   & = &
         \sum\limits_{(j_1,\ldots,j_n) \in J^n}
            p_n(j_1,\ldots,j_n) D(j_1)\otimes D(j_2)\otimes\cdots\otimes D(j_n)
                  \nonumber    . \\
\end{eqnarray}

Then $\{D_n\}$ is $D$-chaotic.

\end{lem}

\proof
 Since the series defining $D_n$ converges in trace norm and
the partial trace operator $\TR^{(n-k)}$ is a bounded operator
with respect to the trace norm, it follows that
\begin{eqnarray*}
     \TR^{(n-k)}D_n & = &
      \sum\limits_{J^n}
            p_n(j_1,\ldots,j_n)
         D(j_1)\otimes D(j_2)\otimes\cdots\otimes D(j_k) \\
     & = &
      \sum\limits_{J^k}
            p^{(k)}_n(j_1,\ldots,j_k)
         D(j_1)\otimes D(j_2)\otimes\cdots\otimes D(j_k) .\\
\end{eqnarray*}
Now $\{p_n^{(k)}\}$ converges weakly to $p^{\otimes k}$
 as $n$ tends to infinity, for $\{p_n\}$ is $p$-chaotic.
Since a sequence of elements of $\ell^1$ converges in norm if and
only if it converges weakly \cite{Banach}, the sequence
$\{p_n^{(k)}\}$ converges in the $\ell^1$ norm to $p^{\otimes k}$
as $n$ tends to infinity. Hence, in trace norm,
\begin{eqnarray*}
          \lim_{n\rightarrow \infty}
           \TR^{(n-k)}D_n        & = &
         \sum\limits_{J^k}
            p(j_1)p(j_2)\cdots p(j_k) \
         D(j_1)\otimes D(j_2)\otimes\cdots\otimes D(j_k)    \\
             & = &
                \bar{D}^{\otimes k},
\end{eqnarray*}
proving that $\{D_n\}$ is $\bar{D}$-chaotic. \hfill $\square$

\subsection{Putting it together: encoding, developing, and reading}

We now formulate two abstract propositions that follow from the
above two lemmas on the encoding procedure.
 Proposition~\ref{Hooray} relies on
Lemma~\ref{lemma1} and is therefore limited by the hypothesis that
the mediating quantum system is finite-dimensional.
Proposition~\ref{Yay} is derived from Lemma~\ref{lemma2}. It
allows the quantum dynamics to take place in an arbitrary Hilbert
space but requires the measurable spaces involved to be discrete.
Despite these technical restrictions there remains a rich variety
of classical examples of the propagation of chaos residing within
each instance of the propagation of quantum molecular chaos.

\begin{prp}
\label{Hooray} Let $\Omega$ be a separable metric space with Borel
$\sigma$-field $\sigma(\Omega)$, let $D(s)$ be a continuous
function from $\Omega$ to the density operators on $\CC^d$, and
let $X:\sigma(\Omega) \longrightarrow \B(\CC^d)$ be a POVM on
$\Omega$. For each $n$, let $\phi_n$ be a normal completely
positive unital endomorphism of $\B((\CC^d)^{\otimes n})$ that
satisfies (\ref{PermutationsPhi}).

 Define the Markov transition kernel $K_n$ on $\Omega^n$ by
\begin{eqnarray*}
    &   & K_n( (\omega_1,\omega_2, \ldots,\omega_n), \ A_1\times A_2 \times \cdots \times A_n)      \\
    &   & \qquad \ = \    \TR\left[ (D(\omega_1)\otimes\cdots \otimes D(\omega_n))
            \    \phi_n(X(A_1)\otimes \cdots \otimes X(A_n))
               \right] .\\
\end{eqnarray*}
If $\{\phi_n\}$ propagates quantum molecular chaos then  $\{K_n\}$
propagates molecular chaos in the classical sense.
\end{prp}

The proof of this proposition is omitted because it follows
directly from Lemma~\ref{propone} and Lemma~\ref{lemma1}.
Likewise, the following proposition follows from
Lemma~\ref{propone} and Lemma~\ref{lemma2}:

\begin{prp}
\label{Yay} Let $\HH$ be a Hilbert space and, for each $n$, let
$\phi_n$ be a completely positive endomorphism of $\B(\HH^{\otimes
n})$ that satisfies (\ref{PermutationsPhi}). Let $J$ be a
countable set equipped with the discrete topology and
$\sigma$-field, and let $X:J \longrightarrow \B(\HH)$ be a POVM on
$J$.

Define the Markov transition matrices $K_n$ on $J^n$ by
\begin{eqnarray*}
    &   &
      K_n((j_1,\ldots,j_n), (j'_1,\ldots,j'_n))       \\
    &   & \qquad \ = \
      \TR \left[(D(j_1)\otimes \cdots \otimes D(j_n))\phi_n(X(j_1')\otimes \cdots \otimes X(j_n'))\right]
             \\
\end{eqnarray*}
If $\{\phi_n\}$ propagates quantum molecular chaos then the
sequence of Markov transition kernels $\{K_n\}$ propagates
classical molecular chaos.
\end{prp}

\subsection{Periodic measurement of complete observables}
\label{complete observables}

The periodic measurement of a complete observable of a quantum
system produces a Markov chain of measurement values.

 Let $\O$ be a complete observable of a
system, represented by a resolution of the identity
$\left\{|e_j\rangle \langle e_j|\right\}_{j \in J}$ for some
orthonormal basis $\{e_j\}_{j\in J}$ of $\HH$.  Consider a
(possibly open) quantum system whose evolution is governed by a
QDS $(\phi)_t$. If the natural quantum evolution of this system is
interrupted periodically by the measurement of $\O$, then the
resulting random sequence of measurement values is a Markov chain
on $J$.   To be specific, suppose the measurements of $\O$ are
performed at times $0, T, 2T, 3T,\ldots$ but there is no other
interference with the evolution $(\phi)_t$.    The first
measurement of $\O$ results in a random outcome, namely, the pure
state
\[
               P_{e_j} \ = \ |e_j\rangle \langle e_j|
\]
into which the system has collapsed.  (We use Dirac notation
$|e\rangle\langle e|$ for projection onto the span of $e$.)  In
effect, measuring the observable $\O$ prepares a pure state
$P_{e_j}$ and informs us of the index $j\in J$ of that pure state.
Having been prepared in the state $P_{e_j}$, the system is allowed
to evolve $T$ time units under $(\phi)_t$.  By time $T$, the state
of the system is $\phi_{T*}(P_{e_j})$.  When $\O$ is measured at
time $T$, the measurement produces another random $j' \in J$, and
the system collapses into the corresponding pure state
$P_{e_{j'}}$. The probability of the transition $j \rightarrow j'$
is
\[
         \TR(\phi_{T*}(P_{e_j})P_{e_{j'}}) \ = \  \TR(P_{e_j}
         \phi_T(P_{e_{j'}})) \ \equiv \ K(j,j'),
\]
defining a Markov transition $K(\cdot,\cdot)$ from $J$ to itself.
At time $T$ the system has been prepared in some pure state
$P_{e_{j'}}$, and a new experiment begins: the system undergoes
$T$ time units of the evolution $(\phi)_t$, transforming its state
from $P_{e_{j'}}$ to $\phi_{T*}(P_{e_{j'}})$, and then $\O$ is
measured at time $2T$, instantaneously forcing the system into the
random state indexed by $j''$ with probability $K(j',j'')$.  This
is repeated, producing a random record of the indices
$j_0,j_1,j_2,\ldots$ of the pure states the system was in upon
measurement at times $0,1,2,\ldots$ of $\O$.  Ideally, these
successive measurement/experiments would be independent, both
physically and stochastically, and it is evident that the
performance of those measurements in succession would produce a
random sequence $j_0,j_1,j_2,\ldots$ governed by the (one-step)
Markov transition kernel $K(\cdot,\cdot)$.

It will be convenient to denote by $ K[(\phi)_t,\O,T]$ the Markov
transition produced in this way, so that
\begin{equation}
\label{derivedTransition}
            K[(\phi)_t,\O,T](j,j') \ = \  \TR(P_{e_j}
         \phi_T(P_{e_{j'}})).
\end{equation}

  Imagine an $n$-component system whose (distinguishable)
components are each quantum systems with the (same) Hilbert space
$\HH$, and which is governed by a QDS $(\phi_n)_t$ that satisfies
the permutation condition (\ref{PermutationsPhi}). Let $\{e_j\}$
be an orthonormal basis for $\HH$ indexed by $J$, and let $\O$ be
the observable that returns the value $j$ if the (component)
system is in the pure state $e_j$.  The observable $\O$ determines
a resolution of the identity $\left\{|e_j\rangle \langle
e_j|\right\}$.  The Hilbert space for the $n$-component system is
$\HH^{\otimes n}$ and the state of the system is a density
operator on $\HH^{\otimes n}$. Let $\O_i$ denote the observable
that returns the value $j$ if the $i^{th}$ component in the pure
state $e_j$. We can imagine measuring $\O_i$ of each of the
components because the components are distinguishable.
Simultaneous measurement of $\O_1,\ldots,\O_n$ on the
$n$-component system results in a random vector
$(j'_1,\ldots,j'_n) \in J^n$ of measurement values, and forces the
system into the pure state $e_{j'_1}\otimes\cdots\otimes
e_{j'_n}$.  Let $ \O^n$ denote the joint measurement
$(\O_1,\ldots,\O_n)$.  Periodic measurement of $\O^n$ results in a
Markov chain of values in $J^n$, since $\O^n$ is a complete
observable of the $n$-component system.  The one-step transition
kernel for this Markov chain is
\begin{equation}
\label{derivedTransition2}
            K[(\phi_n)_t,\O^n,T]({\bf j},\ {\bf j'}) \ = \  \TR(P_{e_{j_1}\otimes\cdots\otimes e_{j_n}}
         \phi_T(P_{e_{j'_1}\otimes\cdots\otimes e_{j'_n}})).
\end{equation}
by formula (\ref{derivedTransition}).

\begin{cor}
\label{interesting}     Let $(\phi_n)_t$ and $\O^n$ be as above
for each $n$, and suppose that $\{(\phi_n)_t\}$ propagates quantum
molecular chaos. Then, for each $T \ge 0$, the sequence of Markov
transitions $\left\{K[(\phi_n)_t, \O^n, T]\right\}_{n\in \NN}$
propagates chaos in the classical sense.
\end{cor}

\proof
 Consider the special case of Proposition~\ref{Yay} where
\[
         D(j) \ = \ X(j) \ = \ |e_j\rangle\langle e_j|.
\]
In that case the sequence $\{L_n\}$ of Markov transitions
\begin{equation}
\label{EllEnn}
      L_n((j_1,\ldots,j_n), (j'_1,\ldots,j'_n))       \ = \
      \TR \left[(Q_{j_1,\ldots,j_n} )
                  \phi_n( Q_{j'_1,\ldots,j'_n} )\right]
\end{equation}
with $ Q_{j_1,\ldots,j_n} = |e_{j_1}\rangle\langle e_{j_1}|
\otimes \cdots \otimes
                  |e_{j_n}\rangle\langle e_{j_n}|  $
propagates molecular chaos.   But comparing (\ref{EllEnn}) to
(\ref{derivedTransition2}) and noting that
\begin{eqnarray*}
         P_{e_{j_1}\otimes\cdots\otimes e_{j_n}}
         & = &
        \big| e_{j_1} \otimes \cdots \otimes e_{j_n} \big\rangle
         \big\langle e_{j_1} \otimes \cdots \otimes e_{j_n} \big|
         \\
          & = &
          |e_{j_1}\rangle\langle e_{j_1}|
                  \otimes |e_{j_2}\rangle\langle e_{j_2}| \cdots \otimes
                  |e_{j_n}\rangle\langle e_{j_n}|
          \ = \
          Q_{j_1,\ldots,j_n}
\end{eqnarray*}
 shows that
\[
       L_n \ = \ K[(\phi_n)_t, \O^n, T].
\]
\hfill $\square$

\section{Acknowledgements}

I am indebted to William Arveson, Sante Gnerre, Lucien Le Cam,
Marc Rieffel, and Geoffrey Sewell for their advice and for their
interest in this research. I would also like to thank Rolando
Rebolledo for his collaboration and support. The author is
supported by the Austrian START project ``Nonlinear Schr\"odinger
and quantum Boltzmann equations."

\end{document}